\documentclass[prl,twocolumn,showpacs,nofootinbib]{revtex4}

\usepackage{graphicx}
\usepackage{amsmath,amsfonts}
\usepackage{dcolumn}
\usepackage{hyperref}

\mathchardef\minus = "002D

\begin{document}

\title{Purely imaginary quasinormal modes of the {K}err geometry}

\author{Gregory B. Cook}\email{cookgb@wfu.edu}
\affiliation{Department of Physics, Wake Forest University,
		 Winston-Salem, North Carolina 27109}
\author{Maxim Zalutskiy}\email{zalump8@wfu.edu}
\affiliation{Department of Physics, Wake Forest University,
		 Winston-Salem, North Carolina 27109}

\date{\today}

\begin{abstract}
We present a method for determining the purely imaginary quasinormal
modes of the Kerr geometry.  Such modes have previously been explored,
but we show that prior results are incorrect.  The method we present,
based on the theory of Heun polynomials, is very general and can be
applied to a broad class of problems, making it potentially useful to
all branches of physics.  Furthermore, our application provides an
example where the method of matched asymptotic expansions seems to
have failed.  A deeper understanding of why it fails in this case may
provide useful insights for other situations.

\end{abstract}


\maketitle

\section{Introduction}
\label{sec:introduction}
The term Quasinormal Mode (QNM) is often used to describe a natural
resonant vibration of an intrinsically dissipative system.  While the
terminology may vary\footnote{For example in biophysics they are
  called quasiharmonic modes.}, the idea is relevant to virtually all
branches of physics (cf
Refs\cite{Settimi-2009,ge-huges-2014,pan-et-al-2015,tourneir-smith-2003}).
The fundamental idea is that, while waves may leave the system, they
may not enter the system.  For black holes in asymptotically flat
spaces, waves may leave the system by flowing into the black hole or
by radiating to infinity.  Represented by a complex frequency
$\omega$, the mode's real part corresponds to the angular frequency
and the imaginary part to the decay rate.  The formalism used to
understand these modes is straightforward\cite{kokkotas-schmidt-1999},
but subtleties can occur.

The existence and nature of gravitational QNMs of the Kerr geometry on
the Negative Imaginary Axis (NIA) have been poorly understood for
quite some time.\footnote{QNM on the NIA are better understood in
  anti-de Sitter spaces (cf Ref.\cite{miranda-zanchin-2006}).  They
  are even known to play a role in the anti-de Sitter/conformal field
  theory correspondence\cite{witczak-krempa-sachdev-2013}.} Early
numerical studies lacked the accuracy necessary to study QNMs near the
NIA\cite{leaver-1985,onozawa-1997}, but general arguments suggested
that QNMs might not exist on the NIA\cite{onozawa-1997}.  The studies
published to date have found modes near the NIA in the neighborhood of
the algebraically special modes of Schwarzschild\cite{chandra-1984} at
$\bar\omega\equiv M\omega=-2i, -10i,
-30i,\ldots$\cite{leaver-1985,onozawa-1997,berticardoso-2003}. And
recent, high-accuracy numerical work\cite{cook-zalutskiy-2014} has
shown that some of these QNMs exist arbitrarily close to the NIA near
$\bar\omega=-2i$. Maassen van den Brink\cite{van_den_brink-2000}
showed that the $s=-2$ algebraically special modes of Schwarzschild
are simultaneously QNMs and left Total Transmission Modes (TTM${}_{\rm
  L}$s)\footnote{For TTM${}_{\rm L}$s, the boundary condition at the
  black hole is reversed, only allowing waves to flow out of the black
  hole.  Alternatively, reversing the boundary condition at infinity
  yields a ``right'' TTM (TTM${}_{\rm R}$).  Reversing both conditions
  yields a bound state.}, finally proving that, at least in the zero
angular momentum limit (ie.\ the Schwarzschild limit), a set of QNMs
do exist precisely on the NIA.  More recently, analytic and numerical
work by Yang {\it et al}\cite{Yang-et-al-2013b} has suggested that a
continuum of purely imaginary QNMs exist for polar modes ($m=0$) near
the extremal limit of the angular momentum $\bar{a}\equiv a/M\to1$.
Inspired by this work, Hod\cite{Hod-2013} derived a related, but more
tailored analytic description of these modes that also imposed the
small frequency limit $|\bar\omega|\ll1$.

\begin{figure}
\includegraphics[width=\linewidth,clip]{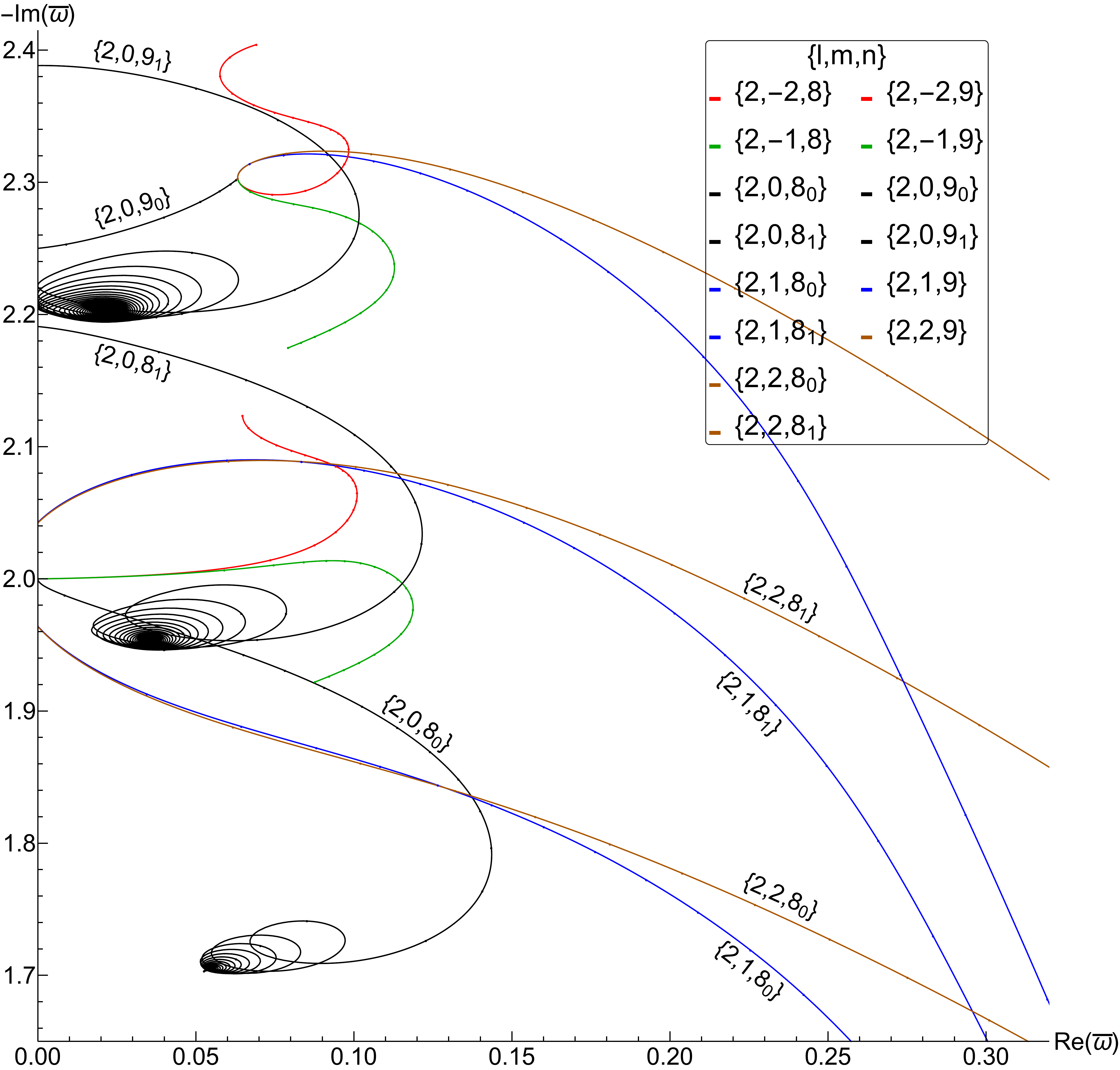}
\caption{\label{fig:qnmplot} The $n=8$ and 9 overtones of the $\ell=2$
  gravitational ($s=-2$) modes of the Kerr geometry. The curves are
  parameterized by $\bar{a}$.  Note that for $n=8$, the $m=1$ and 2
  modes are both multiplets that approach arbitrarily close to the
  NIA. The $m=0$ modes of both overtones are also multiplets, and we
  show four cases where these modes terminate {\em on} the NIA.  The
  $\{2,0,9_1\}$ mode becomes tangent to the NIA at 7 locations.  None
  of these 7 points are QNMs.}
\end{figure}

Using our high-accuracy code for constructing sequences of
QNMs\cite{cook-zalutskiy-2014} of Kerr, parameterized by the angular
momentum $\bar{a}$, we have constructed many full sequences with much
higher overtones than have been previously explored.  In doing this we
have found numerous additional examples of sequences that approach the
NIA {\em at non-vanishing values of $\bar{a}$}.  In fact, for polar
modes, we have found numerous sequences that approach the NIA not
simply once, but thousands of times.  These occur when a sequence
exhibits a looping behavior where the path of the QNMs becomes tangent
to the NIA (see Fig.~\ref{fig:qnmplot} for an example where loops
touch the NIA 7 times).  We have studied all of these cases with high
accuracy and precision.

\begin{figure}
\includegraphics[width=\linewidth,clip]{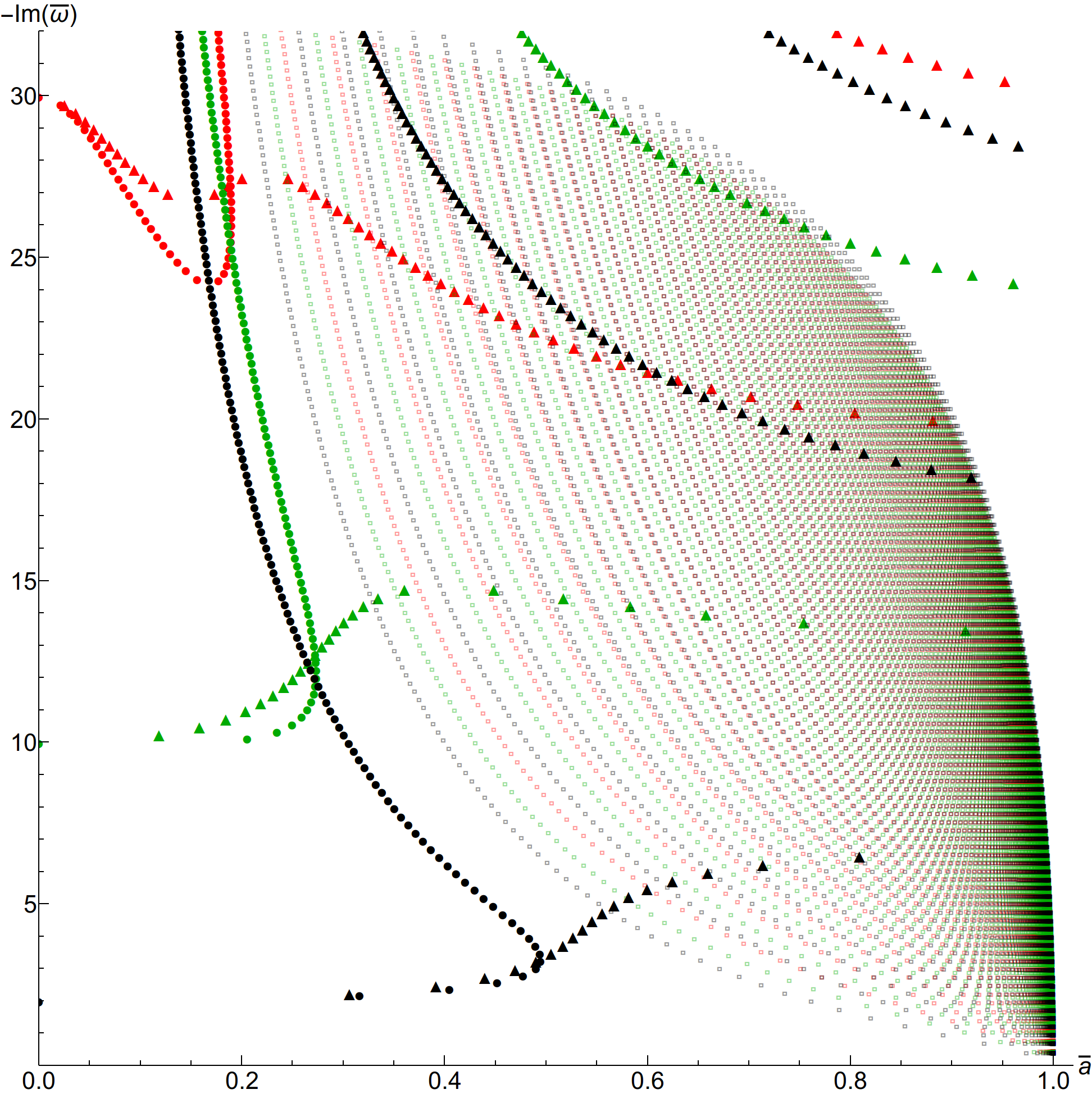}
\caption{\label{fig:m0_l234plot} Polynomial solutions for $\ell=2$, 3,
  and 4.  The solid circles are ``anomalous'' solutions, and so are
  simultaneously QNMs and TTM${}_L$s.  The solid triangles are QNMs.
  And the faint open squares are ``miraculous'' solutions that do not
  represent any kind of mode.  $\ell=2$ modes are in black, $\ell=3$
  in green, and $\ell=4$ in red.  Note the horizontal axis is
  $\bar{a}$, not ${\rm Re}(\bar\omega)$}
\end{figure}

However, we do not find QNMs on or near the NIA in the vicinity of the
numerical results reported by Yang {\it et al}, nor do we find results
that are in agreement with the analytic work of Hod.  To understand
these discrepancies, and to fully understand the nature of QNMs on the
NIA, we have made a careful study of the behavior of modes of Kerr on
the NIA making use of the theory of Heun polynomials\cite{Heun-eqn}.
We find two key results.  First, we find that QNMs can exist on the
NIA {\em only if they correspond to polynomial solutions}.
Furthermore, $m\ne0$ QNMs can only exist on the NIA if they have
frequency $\bar\omega=\bar\omega_\minus$ (see
Eq.~(\ref{eq:Omega_on_NIA})).  We have found no examples with $m\ne0$
that have a QNM {\em on} the NIA.  For example, in
Fig.~\ref{fig:qnmplot} the $\{2,1,8_{0,1}\}$ and $\{2,2,8_{0,1}\}$
sequences all terminate arbitrarily close to, {\em but not on}, the
NIA at a non-vanishing value of $\bar{a}$.  Second, we find that there
are three classes of polynomial solutions on the NIA, each a countably
infinite set.  One class is part of the known extension of the
algebraically special modes of Schwarzschild to cases of non-vanishing
$\bar{a}$\cite{chandra-1984}.  These modes are plotted in
Fig.~\ref{fig:m0_l234plot} as solid circles.  The first 2 solid black
circles correspond to the terminations of $\{2,0,8_0\}$ and
$\{2,0,9_0\}$ {\em on} the NIA in Fig.~\ref{fig:qnmplot}.  Each of
these modes is simultaneously a QNM and a TTM${}_{\rm L}$.  In fact,
this class of solutions adds a previously unknown branch to the
algebraically special modes of Kerr.  The second class of modes are a
previously unknown set of QNMs of Kerr.  These modes are plotted in
Fig.~\ref{fig:m0_l234plot} as solid triangles.  The first 2 solid
black triangles correspond to the terminations of $\{2,0,8_1\}$ and
$\{2,0,9_1\}$ {\em on} the NIA in Fig.~\ref{fig:qnmplot}.  The third
class of modes at first glance appear to be QNMs, but on careful
inspection we find that they exhibit both incoming and outgoing waves
at the event horizon and are neither QNMs nor TTM${}_{\rm L}$s.  These
modes are plotted in Fig.~\ref{fig:m0_l234plot} as faint open squares.
Each looping sequence of QNMs that becomes tangent to the NIA (see
Fig.~\ref{fig:qnmplot}), touches the axis at exactly one of these
frequencies.  However, there are an infinite number of this third
class of modes that do not correspond to a point on any QNM sequence.


\section{Modes on the NIA}
\label{sec:modes_on_NIA}

The Teukolsky master equations describe the dynamics of a massless
field ${}_s\psi$ of spin-weight $s$ in the Kerr geometry,
\begin{equation}\label{eq:Teukolsky_separation_form}
  {}_s\psi(t,r,\theta,\phi) = e^{-i\omega{t}} e^{im\phi}S(\theta)R(r).
\end{equation}
The angular and radial functions, $S(\theta)$ and $R(r)$ respectively,
are obtained by solving a coupled pair of equations, each of which is
of the {\em confluent Heun} form.  Written in {\em nonsymmetrical
  canonical form}, the confluent Heun equation reads
\begin{equation}\label{eq:Heun_NSCF}
\frac{d^2H(z)}{dz^2} 
+ \left(4p + \frac\gamma{z}+\frac\delta{z-1}\right)\frac{dH(z)}{dz}
+ \frac{4\alpha pz -\sigma}{z(z-1)}H(z)=0,
\end{equation}
with regular singular points at $z=0,1$ and an irregular singular
point at $z=\infty$.  Frobenius solutions local to each of the three
singular points can be defined in terms of two functions\cite{Heun-eqn},
\begin{align}
\label{eq:local_a_sol_series}
 Hc^{(a)}(p,\alpha,\gamma,\delta,\sigma;z)&= \sum_{k=0}^\infty{c^{(a)}_kz^k}, \\
\label{eq:local_r_sol_series}
 Hc^{(r)}(p,\alpha,\gamma,\delta,\sigma;z)&= \sum_{k=0}^\infty{c^{(r)}_kz^{-\alpha-k}}.
\end{align}
We are usually interested in solutions called {\em confluent Heun
  functions} which are simultaneously Frobenius solutions for two
adjacent singular points.  However, in this work, polynomial solutions
will play an important role.  {\em Confluent Heun polynomials} are 
simultaneously Frobenius solutions of all three singular points.

The radial Teukolsky equation can be placed into nonsymmetrical
canonical form by making the transformation
{\setlength{\abovedisplayskip}{0pt}\setlength{\belowdisplayskip}{0pt}
\begin{equation}
R(z) = z^\eta (z-1)^\xi e^{(r_+-r_\minus)\zeta z}H(z),
\end{equation}}\noindent
where $r_\pm$ are the radii of the event and Cauchy horizons and
$z\equiv\frac{r-r_\minus}{r_+-r_\minus}$. See
Ref.\cite{cook-zalutskiy-2014,Fiziev-2009b} for a complete
description.  The parameters $\xi$, $\eta$, and $\zeta$ can each take
on one of two values allowing for a total of eight ways to achieve
nonsymmetrical canonical form.  We will focus on the particular cases
where
\begin{equation}\label{eq:Teukolsky_Heun_parameters}
\begin{array}{r}
  \zeta= i\omega\equiv\zeta_+, \\
  \xi=-s -i\sigma_+ \equiv\xi_\minus,
\end{array}
\quad\mbox{with}\quad
 \sigma_+\equiv\frac{2\bar\omega r_+ - ma}{r_+-r_\minus}.
\end{equation}
The choice of $\xi=\xi_\minus$ means that the local solution at the
horizon ($z=1$) given by
\begin{equation}
\label{eq:local_sol_z1a}
  Hc^{(a)}(-p,\alpha,\delta,\gamma,\sigma-4p\alpha;1-z)
\end{equation}
represents waves traveling into the black hole, while $\zeta=\zeta_+$
means the local solution given by Eq.~(\ref{eq:local_r_sol_series})
represents waves traveling out at $z\to\infty$.  These are the boundary
conditions appropriate for QNMs.  The two possible choices for
$\eta=\eta_\pm$ correspond to two possible local behaviors at the
Cauchy horizon $z=0$ and are relevant to the polynomial solutions
below.

\subsubsection{Solutions as confluent Heun functions}
The majority of QNM solutions are found in the form of {\em confluent
  Heun functions}.  These are readily found using ``Leaver's
method''\cite{leaver-1985,leaver-1986} which consists of removing the
asymptotic behavior via $H(z)=z^{-\alpha}\bar{R}(z)$ and rescaling the
radial coordinate as $z\to\frac{z-1}{z}$ so the relevant domain is
$0\le z\le1$.  The solution is expanded as
$\bar{R}(z)=\sum_{n=0}^\infty{a_nz^n}$.  The series has a radius of
convergence of one and more precisely, with
$r_n\equiv\frac{a_{n+1}}{a_n}$,
{\setlength{\abovedisplayskip}{4pt}\setlength{\belowdisplayskip}{0pt}
\begin{equation}\label{eqn:a_ratio_expansion}
   \lim_{n\to\infty} r_n = 1 + \frac{u_1}{\sqrt{n}}
      + \frac{u_2}{n} + \frac{u_3}{n^{\frac32}} +\cdots,
\end{equation}}
and\vspace*{-8pt}
{\setlength{\abovedisplayskip}{2pt}\setlength{\belowdisplayskip}{0pt}
\begin{equation}\label{eq:asymptotic_a}
  \lim_{n\to\infty}a_n \propto n^{u_2} e^{2u_1\sqrt{n}}.
\vspace*{-4pt}
\end{equation}}\noindent
There will be two independent series solutions to the recurrence
relation.  The QNM solutions we seek will be a {\em minimal} solution
we denote by $a_n\to f_n$, and we label the other set of coefficients
by $a_n\to g_n$.  A minimal solution has the property that
$\lim_{n\to\infty}{\frac{f_n}{g_n}}=0$.  With
$u_1=\pm\sqrt{-2i(r_+-r_\minus)\omega}$, the sign choice corresponds
to the two possible asymptotic behaviors for $a_n$.  So long as ${\rm
  Re}(\omega)\ne0$ a minimal solution may exist\footnote{Note that the
  discussion of the sign choice for $u_1$ in
  Ref.~\cite{cook-zalutskiy-2014} contains an error.}.  The ratio
$r_n$ can be written as a continued fraction in terms of the
coefficients of the recurrence relation for the $a_n$.  The key
property of this recurrence relation is given by Pincherle's
theorem\cite{Gautschi-1967} which states that the continued fraction
$r_0$ converges {\em if and only if} the series expansion has a
minimal solution $a_n=f_n$ with $f_0\ne0$.  The continued fraction
$r_0$ must converge to a specific value, and so QNM solutions are
found at those frequencies $\omega$ where $r_0$ does converge to the
required value.

But, it is very important to recognize that for $\omega$ on the NIA,
$\lim_{n\to\infty}\frac{f_n}{g_n}$ becomes oscillatory and {\em a
  minimal solution cannot exist unless the infinite series solution
  terminates}.  Thus, any QNM solution on the NIA must be of the form
of a {\em confluent Heun polynomial}.  Furthermore, {\em the continued
fraction cannot be used to determine the QNM frequencies $\bar\omega$
on the NIA}.

\subsubsection{Polynomial solutions}
\label{sec:Heun_polynomials}
Because confluent Heun polynomials are simultaneous Frobenius
solutions of all three singular points, they can be represented in
several different ways.  To avoid confusion, consider the local
solution around the event horizon at $z=1$ given by
Eq.~(\ref{eq:local_sol_z1a}).  The local Heun solution
$Hc^{(a)}(p,\alpha,\gamma,\delta,\sigma;z)$ is defined by the
three-term recurrence relation
{\setlength{\abovedisplayskip}{2pt}\setlength{\belowdisplayskip}{6pt}
\begin{subequations}
\label{eq:local_a_defs}
\begin{align}
\label{eq:local_a_3term}
  0 =& f^{(a)}_kc^{(a)}_{k+1}+ g^{(a)}_kc^{(a)}_k\!+ h^{(a)}_kc^{(a)}_{k-1}\ :
   \begin{array}{l}c^{(a)}_{-1}=0,\\ c^{(a)}_0=1,\end{array} \\
\label{eq:local_a_g}
  g^{(a)}_k =& k(k-4p+\gamma+\delta-1)-\sigma, \\
\label{eq:local_a_f}
  f^{(a)}_k =& -(k+1)(k+\gamma), \\
\label{eq:local_a_h}
  h^{(a)}_k =& 4p(k+\alpha-1).
\end{align}
\end{subequations}}\noindent
For the series to terminate with order $q$ (a non-negative integer),
two conditions must be satisfied\cite{Heun-eqn}.  First, the lower
diagonal element $h^{(a)}_{q+1}$ must vanish which requires
$\alpha=-q$.  The second condition, denoted by $\Delta_{q+1}=0$, is
that the determinant of the $(q+1)\times(q+1)$ matrix of recurrence
coefficients must vanish:
\begin{equation}\label{eq:Delta_q_cond}
  \left|\begin{array}{ccccc}
  g_0 & f_0 & 0 & \cdots & 0 \\
  h_1 & g_1 & f_1 & \ddots & 0 \\
  0 & h_2 & g_2 & \ddots & 0 \\
  0 & 0 & \ddots & \ddots & \ddots \\
  0 & 0 & \cdots & h_q & g_q
  \end{array}\right|=0.
\end{equation}

The first necessary condition for a polynomial solution to exist,
$\alpha=-q$, can be satisfied in one of two ways\footnote{There are 4
  possible sets of such conditions depending on the choices made for
  the parameters $\xi$ and $\zeta$.  The choice considered here is
  appropriate for considering QNMs.}:
\begin{align}\label{eq:Omega_on_NIA}
\bar\omega = \bar\omega_+\! \equiv
\frac{\bar{a}m-iN_+\sqrt{1-\bar{a}^2}}{2(1+\sqrt{1-\bar{a}^2})}
\ \ \mbox{or}\ \ 
\bar\omega = \bar\omega_\minus\! \equiv -i\frac{N_\minus}{4},
\end{align}
where $N_\pm$ are integers.  These two possibilities are associated
with the two independent local behaviors at the Cauchy horizon
($z=0$).  Choosing either family of possible solutions and fixing a
value for $N_\pm$, a sufficient condition for the existence of a
polynomial solution is to find a root of the $\Delta_{q+1}=0$
condition considered as a function of the angular momentum
$\bar{a}$.\footnote{Each evaluation of the determinant requires a
  solution of the angular Teukolsky equation which is accomplished by
  the spectral method described in Ref.~\cite{cook-zalutskiy-2014}.}
There is, however, one very important caveat.

We must be careful to examine, for each regular singular point, the
behavior of the roots of the indicial equation.  At the event horizon
($z=1$), the two roots are $0$ and $1-\delta$.  Frobenius theory tells
us that, for regular singular points, if the roots of the indicial
equation differ by an integer or zero, in general only one series
solution exists corresponding to the larger root.  A similar condition
holds for irregular singular points, but is not important for this
problem\footnote{The behavior at $z=0$ is governed by $\gamma$ and is
  important when $\bar{a}=0$.  See Ref.\cite{cook-zalutskiy-2014}}.

For the family of solutions associated with $\bar\omega_\minus$ we
have encountered no instances where $\delta$ is an integer.  This
means that polynomial solutions of this family are QNMs.  However the
$\bar\omega_+$ family of solutions is of a form that {\em guarantees}
$\delta$ is a negative integer.  In this case, we find two
possibilities.  The first is that the only series solution behaves
like $(z-1)^{1-\delta}$ instead of $(z-1)^0$.  A solutions of this kind
is referred to as ``anomalous''\cite{van_den_brink-2000} and
represents a solution that is {\em simultaneously a QNM and a
  TTM${}_L$}.  The other possibility is that the $(z-1)^0$ solution
persists.  Solutions of this kind are referred to as
``miraculous''\cite{van_den_brink-2000} and we will show that they
{\em do not represent a mode of any kind}.  These behaviors are, to
say the least, counter-intuitive.

\subsubsection{Anomalous and miraculous solutions}
\label{sec:Anomalous_Miraculous}
Standard Frobenius theory tells us that when the roots of the indicial
equation differ by an integer, only the local series solution
corresponding to the larger root is guaranteed to exist.  The second
local solution will usually include a $\log$ term.  However, it is
possible for the coefficient multiplying this $\log$ term to vanish.
To understand the implications of these two possibilities for our
problem, it is useful to view them from the perspective of scattering
theory.

Consider the ``anomalous'' case.  Begin with an outgoing mode and then
construct an nth-order Born approximation as a function of $\omega$.
We find starting at a certain order in the series, the denominator
contains a term that vanishes when $\bar\omega=\bar\omega_+$.
Essentially, this means that scattering by the potential is so strong
that the normally dominant behavior of the outgoing wave is
overwhelmed and the outgoing wave has exactly the same local behavior
as the incoming wave.  See Ref.~\cite{van_den_brink-2000} for a more
rigorous discussion of this.

In terms of Heun polynomials, when $-\delta$ is an integer less than
$q=-\alpha$, then the upper-diagonal coefficient $f^{(a)}_{-\delta}$
vanishes.  This means that the determinant of the tri-diagonal matrix,
Eq.~(\ref{eq:Delta_q_cond}), can be written as the product of the
determinants of the two diagonal block elements where the blocks are
split following the $1-\delta$ row and column\cite{ElMikkawy-2004}.
Note that the upper off-diagonal block contains all zeros, and the
lower off-diagonal block contains one non-zero element, $h^{(a)}_{1-\delta}$
which {\em cannot} vanish.  In the case of an ``anomalous'' solution,
it is the determinant of the lower diagonal block that vanishes.  This
means that the first $1-\delta$ expansion coefficients vanish and the
leading-order behavior of the Heun polynomial will be
$(z-1)^{1-\delta}$.

For the ``miraculous'' case, we find that it is the determinant of the
upper diagonal block that vanishes.  Returning momentarily to the
example of the Born approximation of the outgoing mode, we find that
the vanishing denominator is countered by a vanishing numerator and in
the limit this term in the Born series is finite.  Again, see
Ref.~\cite{van_den_brink-2000} for a more rigorous discussion.  This
``miraculous'' coincidence is the same miracle that causes the
coefficient of the $\log$ term to vanish in the context of general
Frobenius theory, and is guaranteed by the vanishing of the
determinant of the upper diagonal block in the case of confluent Heun
polynomials.

The subtlety of the ``miraculous'' case stems from the fact that the
limiting value of the coefficient of the $(z-1)^{1-\delta}$ term in
the expansion of the Born approximation is not the same as the
coefficient of this term in the Heun polynomial solution.  In short,
the local behavior of the Heun polynomial solution at $z=1$ is a
linear combination of the local behaviors of an incoming and an
outgoing wave\footnote{In all of the miraculous cases we have
  examined, neither the incoming nor outgoing mode is polynomial at
  $\bar\omega=\bar\omega_+$.  However, one linear combination causes
  all terms at order $q+1$ and above to vanish.}.  Given the
polynomial solution, we can explicitly construct a linearly
independent solution with the same $\bar{a}$ and $\bar\omega$.  This
solution has the behavior of an incoming wave at $z=1$ potentially
allowing us to subtract away this unwanted contribution from the
polynomial solution.  However, it behaves like an incoming wave at
infinity as well, so we cannot fix the behavior at one boundary
without spoiling it at the other.  Thus, these ``miraculous''
solutions are neither QNMs nor TTM${}_L$s.

\section{Summary and Discussion}

The approach we have described allows for a complete determination of
the QNMs of the Kerr geometry on the NIA.  For the case of
gravitational perturbations $s=-2$ we have compared our results to
QNMs in the neighborhood of the NIA that were obtained with high
numerical accuracy.  In all three classes described above, we find the
limiting behavior of the numerical results to be in precise agreement
with the frequency $\bar\omega$ and angular momentum $\bar{a}$
associated with the polynomial solutions.  The agreement spans more
than a thousand instances and at times required numerical results
computed to an accuracy of $10^{-12}$ or better to accurately
distinguish between possible solutions.

We have shown that there are many solutions that are clearly QNMs on
the NIA ($\bar\omega_\minus$), and many that are simultaneously QNMs
and TTM${}_L$s on the NIA (anomalous cases of $\bar\omega_+$).  In
fact, it seems clear that there should be a countably infinite set of
both such modes.  We have also shown that a third class of polynomial
solutions (miraculous cases of $\bar\omega_+$) are {\em not} QNMs even
though on first inspection they would appear to satisfy the required
criteria.  Additional details on the methods to find these modes and
their interesting behavior will be discussed in a future
paper\cite{cook-zalutskiy-2016b}

The numerical solutions for QNMs on the NIA found in
Ref.~\cite{Yang-et-al-2013b} used Leaver's method for finding QNMs as
the roots of a continued fraction.  While the majority of their
numerical solutions are valid, we have shown that this method cannot
be used to find QNMs on the NIA and their solutions on the NIA are not
valid.  Yet these misidentified QNMs appear in good agreement with the
analytic approximations in Refs.~\cite{Yang-et-al-2013b} and
\cite{Hod-2013}.  First, we note that Hod's analytic description of
these ``QNMs'' (see Eq.~(21) of Ref.~\cite{Hod-2013}) takes a form
identical to our $\bar\omega_+$ condition.  The important difference,
using our notation, is that his modes require $N_+$ be near to, but
{\em explicitly not}, an integer.  Furthermore, he finds a nearby
total reflection mode (a TTM${}_L$ in our terminology).  It is
possible that Hod's solutions result from a ``splitting of the exact
solution'' caused by the various approximations that were made.  It is
also likely that the matched asymptotic expansions used in these
works\cite{Yang-et-al-2013b,Hod-2013} are forcing a general solution
to apply in a situation where only a polynomial solution is allowed.
A deeper understanding of why these analytic methods fail in this
special situation might be very helpful for numerous other situations.

\acknowledgments
We would like to thank Emanuele Berti, Aaron Zimmerman, and Shahar Hod
for helpful discussions.

\end{document}